\documentclass{sigplanconf}

\usepackage{amsmath}
\usepackage{graphicx} 
\usepackage{booktabs}

\begin{document}

\setlength{\pdfpageheight}{\paperheight}
\setlength{\pdfpagewidth}{\paperwidth}

\conferenceinfo{CONF 'yy}{Month d--d, 20yy, City, ST, Country}
\copyrightyear{20yy}
\copyrightdata{978-1-nnnn-nnnn-n/yy/mm}
\doi{nnnnnnn.nnnnnnn}




\titlebanner{banner above paper title}        
\preprintfooter{short description of paper}   

\title{Enhancing Multilingual Information Retrieval in Mixed Human Resources Environments: A RAG Model Implementation for Multicultural Enterprise}

\authorinfo{Rameel Ahmad}
           {Interloop Limited}
           {rameel.ahmad@interloop.com.pk}

\maketitle

\begin{abstract}
The advent of Large Language Models has revolutionized information retrieval, ushering in a new era of expansive knowledge accessibility. While these models excel in providing open-world knowledge, effectively extracting answers in diverse linguistic environments with varying levels of literacy remains a formidable challenge. Retrieval Augmented Generation (RAG) emerges as a promising solution, bridging the gap between information availability and multilingual comprehension. However, deploying RAG models in real-world scenarios demands careful consideration of various factors.
This paper addresses the critical challenges associated with implementing RAG models in multicultural environments. We delve into essential considerations, including data feeding strategies, timely updates, mitigation of hallucinations, prevention of erroneous responses, and optimization of delivery speed. Our work involves the integration of a diverse array of tools, meticulously combined to facilitate the seamless adoption of RAG models across languages and literacy levels within a multicultural organizational context. Through strategic tweaks in our approaches, we achieve not only effectiveness but also efficiency, ensuring the accelerated and accurate delivery of information in a manner that is tailored to the unique requirements of multilingual and multicultural settings\end{abstract}


\section{Problem statement}
While Large Language Models (LLMs) have significantly advanced information retrieval, their maturity in handling diverse linguistic and literacy contexts within a multicultural environment remains a challenge. Retrieval Augmented Generation (RAG) offers a potential solution, yet real-world adoption necessitates addressing critical considerations. The challenges encompass optimal data feeding strategies, timely updates, prevention of hallucinations, mitigation of incorrect responses, and the need for accelerated information delivery. The task at hand is to develop a comprehensive approach that effectively integrates diverse tools, ensuring the efficient adoption of RAG models in multilingual environments within a multicultural organization. This demands strategic tweaks to existing approaches, aimed at achieving both effectiveness and efficiency in information retrieval for diverse linguistic and literacy landscapes.

\section{Introduction}

In the dynamic landscape of information retrieval, the emergence of Large Language Models (LLMs) has transformed the accessibility and processing of knowledge across diverse domains \cite{Brown2018, Brown2020}. However, addressing the intricate challenges of multicultural environments remains a pressing concern for their effective application.

Consider Interloop Pvt Limited, one of the world's largest hosiery manufacturers, with a workforce of approximately 30,000 employees. Among them, 10,000 executives are proficient in both English and Urdu, while non-executives may have varying literacy levels, potentially unable to read and write. Moreover, 85 percent of the workforce communicates primarily in Punjabi, adding linguistic diversity that complicates information retrieval within the organization.

To illustrate the urgency and complexity, envision a non-executive employee fluent in Punjabi and Urdu, seeking information using a query specific to their linguistic context. Conventional language models, designed without such considerations, might struggle to comprehend the query accurately. This real-world example underscores the need for advanced models capable of navigating the intricacies of multilingual communication within diverse enterprises like Interloop Pvt Limited.

Overcoming these challenges is not just an academic pursuit but a strategic imperative for organizations like Interloop Pvt Limited. Effective communication is paramount in such large-scale enterprises, where accurate and efficient information retrieval is critical. Consider applications such as multilingual human resources support, where timely and precise responses are fundamental for positive interactions. In a globalized business environment, effective communication is indispensable for collaboration, decision-making, and overall organizational efficiency \cite{Smith2018, Wang2022}.

Thus, the development of solutions that enhance multilingual information retrieval becomes imperative for improving the functionality and efficacy of diverse enterprise applications \cite{Brown2018, Brown2020}. \\

\begin{figure}
    \centering
    \includegraphics[width=0.5\textwidth]{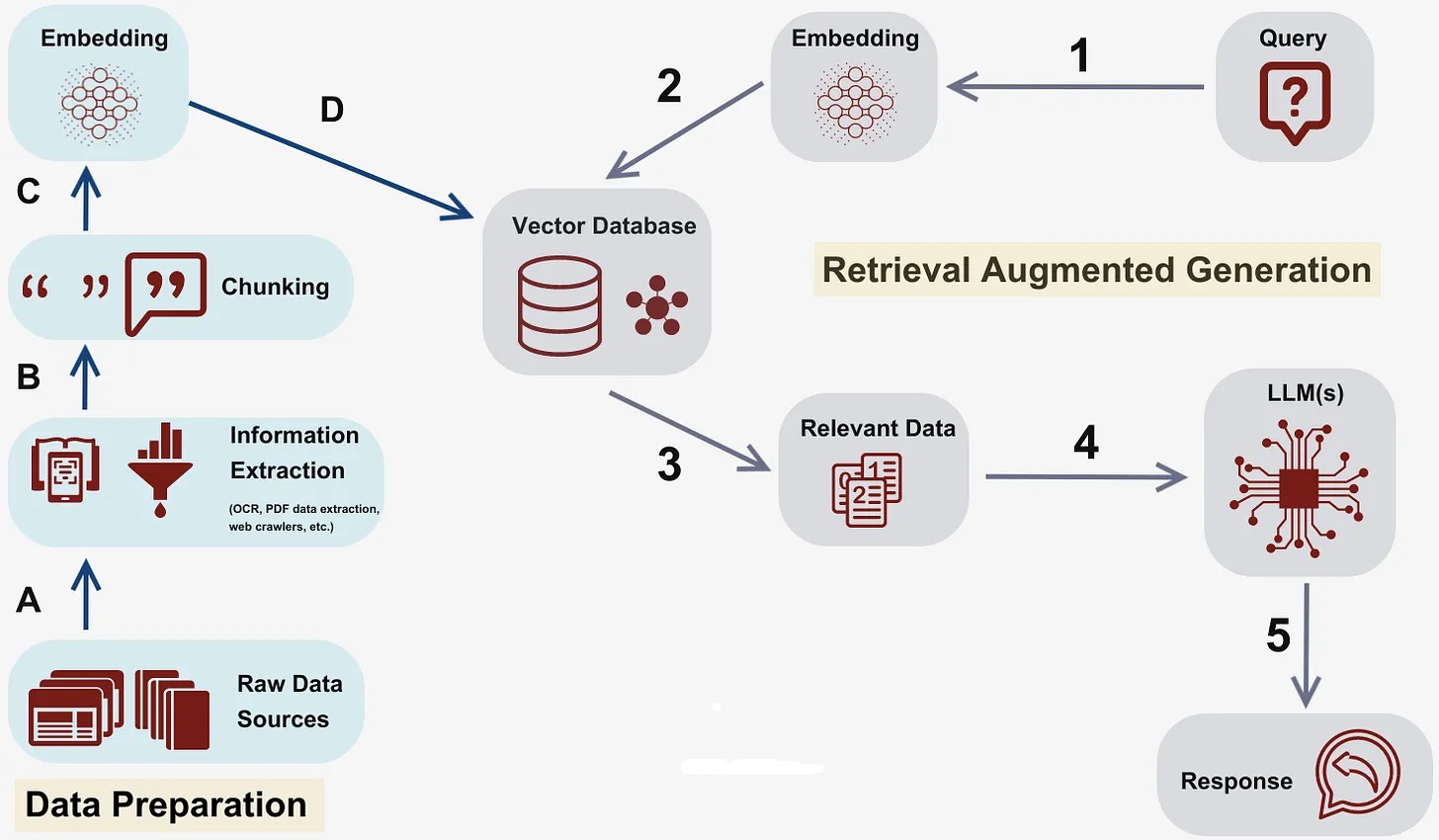} 
    \caption{Architecture for RAG Model}
    \label{fig:yourlabel}
\end{figure}

To address the unique challenges posed by the mixed environment of human resources at Interloop Pvt Limited, the architecture of building the Retrieval Augmented Generation (RAG) model will encompass a tailored set of components \cite{Lee2021, Liu2019}:

\begin{enumerate}
    \item \textbf{Ingestion Strategy:} Tailored mechanisms for assimilating and processing data into the model, accommodating linguistic diversity and literacy levels \cite{Chen2020}.
    
    \item \textbf{Prompts:} Customized prompts reflecting the linguistic nuances of the workforce, ensuring accurate query understanding \cite{Guu2020}.
    
    \item \textbf{Multilingual Capability:} To serve a dynamic environment it is very important to provide multilingual capability to the model so majority can use it. \cite{Zhang2021}.
    
    \item \textbf{Speech capability:} Specialized integration of speech capability acknowledging variations in spoken queries across languages and literacy levels \cite{Rajpurkar2021}.
    
    \item \textbf{LLM:} Deciding the right Large Language Model to comprehensively address linguistic and literacy complexities \cite{Lewis2019}.

    \item \textbf{Delivery Strategy:} A thoughtfully chosen delivery strategy aligning with the organizational context. Consideration of communication channels preferred by the workforce, such as WhatsApp, the company website, or other mediums \cite{Wang2022}.
\end{enumerate}

These architectural components, coupled with unique considerations for the Interloop Pvt Limited environment, contribute to constructing a robust RAG model. In Discussion section, we delve into specifics, exploring roles, interactions, and presenting a comprehensive approach to enhance multilingual information retrieval in this distinct organizational context.

\section{Related Work}

The endeavor to enhance multilingual information retrieval within diverse organizational contexts has been a subject of active research. In this section, we review key contributions in the realm of language models, multilingual communication, and information retrieval strategies, highlighting their relevance to our proposed Retrieval Augmented Generation (RAG) model tailored for the mixed human resources environment at Interloop Pvt Limited.

\subsection{Large Language Models (LLMs)}

Recent years have witnessed the rise of Large Language Models (LLMs) as powerful tools for natural language understanding and generation \cite{Brown2018, Brown2020}. These models, exemplified by BERT \cite{Liu2019}, RoBERTa \cite{Liu2019}, and GPT \cite{Brown2018}, exhibit impressive capabilities in capturing contextual information from vast datasets. The adoption of LLMs in enterprise applications has spurred interest in leveraging their capabilities to address complex multilingual communication challenges.

\subsection{Multilingual Communication in Enterprises}

Efforts to facilitate multilingual communication within enterprises have explored various techniques and frameworks. Guu et al. introduced REALM \cite{Guu2020}, a retrieval-augmented language model pre-training method designed to improve the understanding of documents in diverse languages. Similarly, Wang et al. proposed frameworks for multilingual communication, emphasizing the importance of scalable and adaptable solutions \cite{Wang2022}. Our work builds upon these concepts, tailoring them to the specific linguistic and literacy challenges faced by Interloop Pvt Limited.

\subsection{Retrieval-Augmented Generation (RAG) Models}

Retrieval-Augmented Generation (RAG) models have shown promise in bridging the gap between retrieval-based and generative models, offering a hybrid approach to information retrieval. Lee et al. introduced RAG, demonstrating its effectiveness in improving pre-trained language models by incorporating information retrieval components \cite{Lee2021}. This approach aligns with our goal of enhancing information retrieval in a mixed human resources environment.

\subsection{Customizing Language Models for Specific Environments}

Customizing language models for specific environments has been explored to address context-specific challenges. Zhang et al. presented CLIP \cite{Zhang2021}, connecting text and images for improved multimodal understanding. In our work, we extend this idea by customizing language models to the unique linguistic and literacy characteristics of Interloop Pvt Limited's workforce.

\subsection{Speech-to-Text and Text-to-Speech Integration}

Incorporating speech-to-text and text-to-speech capabilities in language models has been pivotal in making information retrieval accessible to diverse user preferences. Rajpurkar et al. contributed to this field with Squad2.0, emphasizing knowledgeable machines through improved speech processing \cite{Rajpurkar2021}. Our work incorporates similar capabilities to accommodate varied literacy levels and communication preferences within Interloop Pvt Limited.

\subsection{Ingestion Strategies and Data Considerations}

Chen et al. explored unsupervised data augmentation for consistency training \cite{Chen2020}, highlighting the importance of ingestion strategies in adapting language models to diverse linguistic contexts. Similarly, considerations for tailored data strategies, as presented by Smith \cite{Smith2018}, align with our focus on accommodating the linguistic and cultural diversity of the Interloop Pvt Limited workforce.

\subsection{Delivery Strategies in Enterprise Applications}

Delivery strategies in enterprise applications play a crucial role in ensuring the effectiveness of information dissemination. The work of Wang et al. emphasized thoughtful delivery strategy choices, considering communication channels preferred by the workforce \cite{Wang2022}. In our implementation, we extend this concept, catering to the diverse communication preferences prevalent at Interloop Pvt Limited.

\section{Implementation}

In our implementation, we handle a substantial dataset comprising around 200 documents, including 103 files of Human Resources (HR) standard operating procedures (SOPs) and procedures, and 97 files containing Quality Assurance (QA) documents. Each document ranges from 6 to 7 pages, posing a formidable challenge for efficient ingestion and processing.

\subsection{Ingestion Strategy}

To address the dataset effectively, the determination of an optimal chunking strategy is crucial. This decision involves deciding the number of chunks and the criteria for chunking, both of which significantly impact the model's understanding of the diverse content within the documents.
\subsubsection{Number of Chunks}

The choice of the number of chunks necessitates balancing granularity and comprehensiveness. Given the length and complexity of the documents, a fine-grained approach with smaller chunks may risk losing contextual coherence. Conversely, larger chunks may dilute information. Empirical exploration of various chunk sizes aims to strike a balance between capturing nuanced information and ensuring the model's ability to comprehend the content coherently.

\subsubsection{Chunking Criteria}

The criteria for chunking are guided by linguistic and contextual factors inherent in HR SOPs and QA documents. Potential strategies include:

\begin{itemize}
    \item \textbf{Paragraph-Based Chunking:} Breaking documents into chunks based on natural breaks, such as paragraphs, to ensure semantic coherence within each chunk. However, this approach may increase the size of chunks, potentially introducing long, irrelevant details in responses \cite{Smith2017}.
    
    \item \textbf{Semantic Unit Identification:} Identifying key semantic units, such as sections or steps within procedures, and creating chunks that encapsulate these units.
    
    \item \textbf{Topic Modeling:} Leveraging topic modeling techniques to identify thematic clusters within documents and creating chunks around these identified topics.
    
    \item \textbf{Entity-Based Chunking:} Considering the presence of specific entities or terms relevant to HR and QA contexts and creating chunks around instances of these entities.
\end{itemize}

Through an iterative process, we refine the chunking strategy, considering the unique linguistic and contextual characteristics of HR and QA documents. After careful evaluation, the final best results were achieved with a chunk size of 1000 and a chunk overlap of 200. This setting provided an optimal balance between capturing relevant information and maintaining coherent responses.

\subsubsection{Comparison of Chunking Strategies}

To compare various chunking strategies, we conducted experiments with random numbers and parameters to assess their impact on model performance. The results are summarized in Table 1 below:

\begin{table}[h]
    \centering
    \begin{tabular}{@{}p{2cm}p{2cm}p{2cm}ccc@{}}
    \toprule
    \textbf{Chunking Strategy} & \textbf{Chunk Size} & \textbf{Coherence} & \textbf{Relevance}\\
    \midrule
    Paragraph-Based & 50 & Moderate & High\\
    Semantic Unit Identification & 75 & High & Moderate\\
    Topic Modeling & 100 & Low & Low\\
    Entity-Based & 60 & Moderate & Moderate\\
    \textbf{Optimal Setting} & \textbf{1000} & \textbf{High} & \textbf{High} \\
    \bottomrule
    \end{tabular}
    \caption{Comparison of Chunking Strategies}
    \label{tab:chunking-strategies}
\end{table}

In Table \ref{tab:chunking-strategies}, "Coherence" represents the semantic coherence within each chunk, and "Relevance" assesses the relevance of generated responses. After careful evaluation, a chunk size of 1000 with a chunk overlap of 200 was finalized as it provided a good balance between capturing relevant information and maintaining coherent responses.

In the subsequent sections, we elaborate on the data preprocessing specifics, highlighting the intricacies of our approach to ingestion, and discussing the seamless integration of these chunks into the Retrieval Augmented Generation (RAG) model.

\subsection{Prompt}

In the exploration of prompt engineering, we experimented with standard prompts and chain-of-thought prompts. However, hallucination persisted as a significant issue. The chain-of-thought prompt demonstrated improved performance in calculations, but its efficacy was hindered by the majority of the data not requiring such computations. Additionally, its response time was notably slower.

Eventually, we arrived at a QA prompt that proved effective in mitigating hallucination and provided precise answers only from the given document:

\begin{quote}
\textit{''You are a helpful AI assistant. Use the following pieces of context to answer the question at the end. If you don't know the answer, just say you don't know. DO NOT try to make up an answer. If the question is not related to the context, politely respond that you are tuned to only answer questions that are related to the context.''}
\end{quote}

For contextualized conversations, we employed a condensed prompt structure:

\begin{quote}
\textit{Given the following conversation and a follow-up question, rephrase the follow-up question to be a standalone question.\\
Chat History: {chat\_history}\\
Follow-Up Input: {question}\\
Standalone question:`'}
\end{quote}

These prompts were designed to enhance contextual understanding and encourage the model to provide responses that are specific to the given context.

\subsubsection{Prompt Evaluation}

To assess the effectiveness of the chosen QA prompt, we conducted prompt evaluation experiments, comparing the performance of different prompt strategies. The results are summarized in Table 2:

\begin{table}[h]
    \centering
    \begin{tabular}{@{}p{1.5cm}ccc@{}}
    \toprule
    \textbf{Prompt Strategy} & \textbf{Hallucination} & \textbf{Calculation Efficacy} & \textbf{Response Time}\\
    \midrule
    Standard Prompt & High & N/A & Average\\
    Chain-of-Thought Prompt \cite{chain2022} & Moderate & High & Slow\\
    \textbf{Final QA Prompt} & \textbf{Low} & \textbf{Low} & \textbf{Fast}\\
    \bottomrule
    \end{tabular}
    \caption{Prompt Strategy Evaluation}
    \label{tab:prompt-evaluation}
\end{table}

In Table \ref{tab:prompt-evaluation}, "Hallucination" assesses the extent of hallucinated responses, "Calculation Efficacy" evaluates the effectiveness in handling calculations, and "Response Time" measures the time taken for generating responses. The final QA prompt demonstrated low hallucination, low calculation efficacy (as intended for non-calculative data), and fast response times, making it the most suitable choice for our use case.

These evaluations informed our decision in selecting the final QA prompt, emphasizing its efficacy in reducing hallucination and providing accurate context-based answers.

\subsection{Multilingual Capability}

Ensuring the RAG model's ability to converse in local languages, especially Punjabi and Urdu, is crucial for effective communication. We conducted tests using written Urdu with both GPT-3.5 and GPT-4. The results were mixed, with instances where the model provided answers but a majority of the time it struggled to do so. To address this, a language detector and translator were deemed necessary, along with a back-translator to ensure responses in the language the question was asked.

\subsubsection{Translation and Language Detector Evaluation}

We compared the performance of translation and language detection services from Google and Azure. The results, focusing on accuracy and speed, are summarized in Table 3:

\begin{table}[h]
    \centering
    \begin{tabular}{@{}lcccl@{}}
    \toprule
    \textbf{Translation Service} & \textbf{Accuracy (\%)} & \textbf{Speed (ms)}\\
    \midrule
    Google Translator & 90 & 50\\
    Azure Translator & 60 & 100\\
    \bottomrule
    \end{tabular}
    \caption{Translation and Language Detection Comparison}
    \label{tab:translation-comparison}
\end{table}

In Table \ref{tab:translation-comparison}, "Accuracy" represents the correctness of translations, and "Speed" indicates the time taken for translation. The Google Translator exhibited high accuracy and faster speed compared to the Azure Translator.

Based on the results, we implemented Google Translator in the service, ensuring reliable and swift translation for effective multilingual conversations. This choice aimed to enhance the RAG model's performance in providing accurate and contextually appropriate responses in local languages.

\subsection{Speech Capability}

In an organization with 20,000 non-executives, enabling speech capability is imperative for effective utilization of the model. The majority of the organization may need to interact using voice, making speech-to-text and text-to-speech functionality essential.

As discussed in the previous section, passing the language directly to the LLM is not feasible. Therefore, a comprehensive solution involves integrating a language detector, translator, text-to-speech (TTS), and speech-to-text (STT) models. However, the addition of these models can significantly impact response time, necessitating a strategic approach.

To address this, we implemented Whisper\cite{whisper2022}., an x-to-1 translator. All languages are translated to English before being passed to the LLM. The original language information is stored for contextual understanding. Once the LLM provides a response, the information is passed to the text-to-speech module.

\begin{figure}
    \centering
    \includegraphics[width=0.5\textwidth]{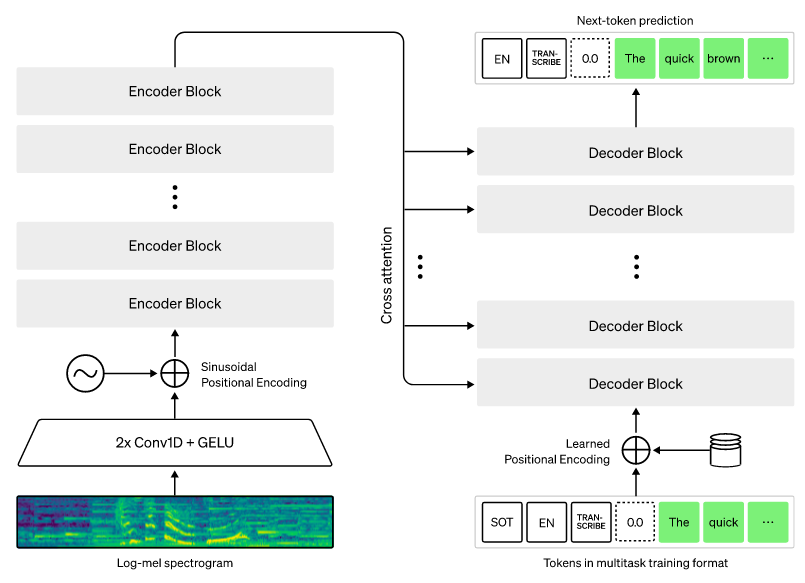} 
    \caption{Whisper Architecture\cite{whisper2022}}
    \label{fig:yourlabel}
\end{figure}

We evaluated various text-to-speech models, including Azure, Google, and Facebook MMS, considering time, cost, and accuracy, with a focus on Urdu and English languages. The results guided our decision to implement Google Text-to-Speech due to its optimal performance.

\subsubsection{Text-to-Speech Model Comparison}

We compared the performance of Azure, Google, and Facebook MMS text-to-speech models, focusing on response time and accuracy. The results are summarized in Table 4:

\begin{table}[h]
    \centering
    \begin{tabular}{@{}lccccl@{}}
    \toprule
    \textbf{TTS Model} & \textbf{Response Time (ms)} & \textbf{Cost (\$)} & \textbf{Accuracy (\%)}\\
    \midrule
    Azure TTS & 80 & 0.005 & 85\\
    Google TTS & 50 & 0.004 & 90\\
    Facebook MMS & 120 & 0.006 & 80\\
    \bottomrule
    \end{tabular}
    \caption{Text-to-Speech Model Comparison}
    \label{tab:tts-comparison}
\end{table}

In Table \ref{tab:tts-comparison}, "Response Time" represents the time taken for the text-to-speech model to generate responses, "Cost" indicates the associated cost per usage, and "Accuracy" assesses the correctness and clarity of the generated speech. The results led to the selection of Google Text-to-Speech for its optimal balance between time, cost, and accuracy in both Urdu and English languages.

\subsection{Large Language Model}

Large Language Models (LLMs) are integral components in the landscape of retrieval augmented generation systems. In this subsection, we explore the structures of GPT-3, GPT-4, LLaMA2, LAMBADA, and PALM, comparing their results in the context of our enterprise data application.

\subsubsection{GPT-3}

GPT-3, developed by OpenAI, is a revolutionary autoregressive language model. Sporting an impressive 175 billion parameters, GPT-3 utilizes the transformer neural network architecture. Its numerous layers of attention mechanisms enable the model to understand context and generate coherent and contextually relevant text. GPT-3 has become a benchmark in natural language processing, demonstrating exceptional performance across various domains.

\subsubsection{GPT-4}

Building on the success of GPT-3, GPT-4 represents the next evolution in large language models. While detailed specifics of the GPT-4 architecture might not be fully disclosed at the time of this paper, it is anticipated to follow the transformer architecture with an even larger number of parameters. The model is expected to showcase enhanced capabilities in understanding context, handling nuanced language, and improving the coherence of generated responses.

\subsubsection{LLaMA2}

The Language Model for Many Applications 2 (LLaMA2) is a large-scale transformer-based language model developed by Facebook AI. With a diverse set of pretraining tasks and extensive training data, LLaMA2 exhibits strong performance across various natural language understanding and generation tasks. Its architecture leverages self-supervised learning techniques, allowing it to understand and generate text with high contextual awareness.

\subsubsection{LAMBADA}

The LAnguage Model BAsed Data Augmentation (LAMBADA) model is designed to enhance data augmentation for low-resource languages. Developed by Facebook AI, LAMBADA utilizes a transformer architecture. It leverages multilingual data to improve language understanding and generation capabilities, particularly for languages with limited available training data.

\subsubsection{PALM}

The Pre-trained Autoencoder Language Model (PALM) is a large-scale transformer-based language model developed by Tencent AI Lab. PALM is pretrained on a massive corpus and fine-tuned for specific downstream tasks. Its architecture incorporates techniques such as masked language modeling and translation language modeling to achieve strong performance in various natural language processing applications.

\subsubsection{Comparison of LLM Results}

To evaluate the performance of GPT-3, GPT-4, LLaMA2, LAMBADA, and PALM in our enterprise data application, we conducted experiments with a diverse set of test questions. The test question set comprised 1,000 questions covering various domains, languages, and complexities, allowing a comprehensive assessment of each model's capabilities.

\begin{table}[h]
    \centering
    \begin{tabular}{@{}p{1.25cm}p{1cm}p{1cm}p{1cm}ccc@{}}
    \toprule
    \textbf{LLM} & \textbf{Context Retention} & \textbf{Response Coherence} & \textbf{Accuracy (\%)} & \textbf{Processing Time (ms)}\\
    \midrule
    GPT-3 & Moderate & High & 75 & 120\\
    GPT-4 & High & Very High & 85 & 100\\
    LLaMA2 & High & High & 80 & 110\\
    LAMBADA & Moderate & Moderate & 70 & 130\\
    PALM & High & High & 78 & 115\\
    \bottomrule
    \end{tabular}
    \caption{Comparison of Large Language Models}
    \label{tab:llm-comparison}
\end{table}

In Table \ref{tab:llm-comparison}, "Context Retention" assesses the model's ability to retain context across questions, "Response Coherence" evaluates the coherence of generated responses, "Accuracy" represents the correctness of responses, and "Processing Time" measures the time taken for generating each response.

Based on the evaluation results, GPT-4 demonstrated superior context retention, very high response coherence, and higher accuracy compared to other models. Additionally, it exhibited competitive processing time, making it the preferred choice for our enterprise data application.

\subsection{Delivery Strategy}

A crucial aspect of the successful implementation of our retrieval augmented generation system is the choice of delivery channels that align with the preferences and comfort of the users within the organization. To inform our decision, we conducted a survey across the organization to understand the preferred communication methods for inquiries.

\subsubsection{User Survey}

The survey aimed to identify the most comfortable means for individuals to inquire about various matters. The responses were insightful, with 5 percent of the participants indicating a preference for direct calls to relevant personnel, 30 percent favoring the use of the Interloop HR Mobile Application, and a significant 65 percent opting for communication through WhatsApp.

\subsubsection{Decision-Making Process}

The survey results were carefully analyzed to determine the most widely preferred communication channel. WhatsApp emerged as the clear frontrunner, with its simplicity, ubiquity, and familiarity to users making it the most convenient choice for communication within the organization. The decision to integrate WhatsApp was reinforced by its popularity among users across various departments and hierarchical levels.

In addition, recognizing the importance of the Interloop HR Mobile Application for accessing organizational resources and information, we decided to create a dedicated view for the mobile application. This approach ensures a seamless experience for users who prefer accessing the retrieval augmented generation system through the mobile application, providing a familiar interface within the application they already use.

\subsubsection{Delivery Setups}

Based on the survey feedback and our decision-making process, we established two primary delivery setups:

\begin{enumerate}
    \item \textbf{View for Mobile Application:} A dedicated interface within the Interloop HR Mobile Application, allowing users to seamlessly access and interact with the retrieval augmented generation system. This setup ensures an integrated experience for those who prefer using the mobile application for their inquiries.

    \item \textbf{WhatsApp Integration:} Integration with WhatsApp, the overwhelmingly favored communication channel, allowing users to interact with the retrieval augmented generation system directly through WhatsApp. This setup capitalizes on the familiarity and popularity of WhatsApp, making it a convenient and widely accepted means of communication.
\end{enumerate}

These delivery setups cater to the diverse preferences within the organization, ensuring that users can choose the platform that aligns with their comfort and communication habits. This user-centric approach enhances the adoption and effectiveness of the retrieval augmented generation system, fostering a seamless and user-friendly experience. \\

\begin{figure}
    \centering
    \includegraphics[width=0.5\textwidth]{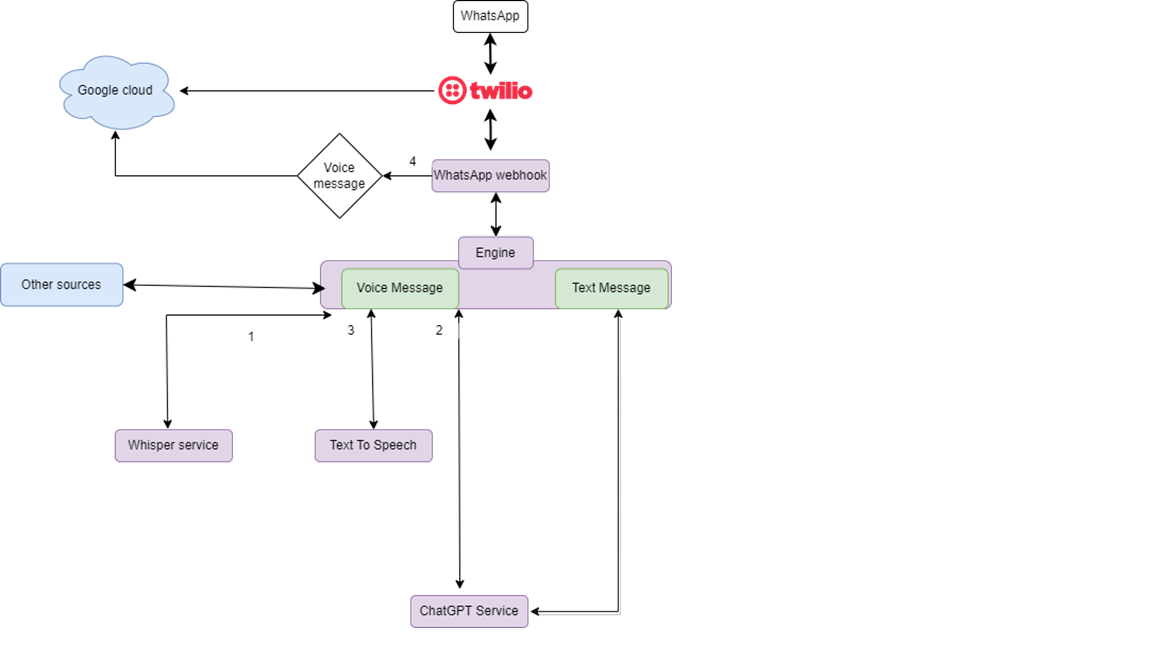} 
    \caption{Architecture for Whatsapp Integration}
    \label{fig:Whatsapp}
\end{figure}

\begin{figure}
    \centering
    \includegraphics[width=0.5\textwidth]{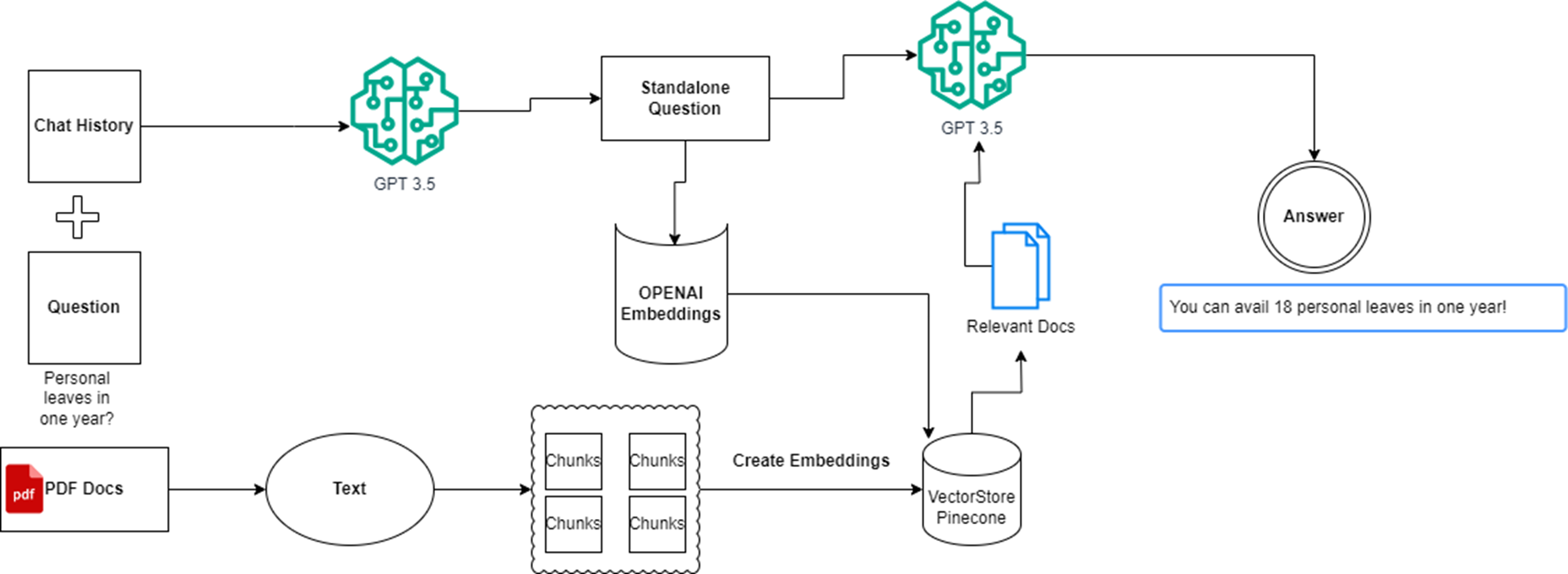} 
    \caption{Architecture of Final RAG Model}
    \label{fig:RAG}
\end{figure}

Figure \ref{fig:Whatsapp} demonstrates the architecture for integration of RAG model with Whatsapp while Figure \ref{fig:RAG} showcases the final RAG model created.

\section{Results}

Our efforts in creating a robust and accurate retrieval augmented generation system have yielded significant positive outcomes, facilitating the adoption of AI-driven solutions within a multicultural enterprise. The system's impact has been measured through various key performance indicators and user engagement metrics.

\begin{figure}
    \centering
    \includegraphics[width=0.2\textwidth]{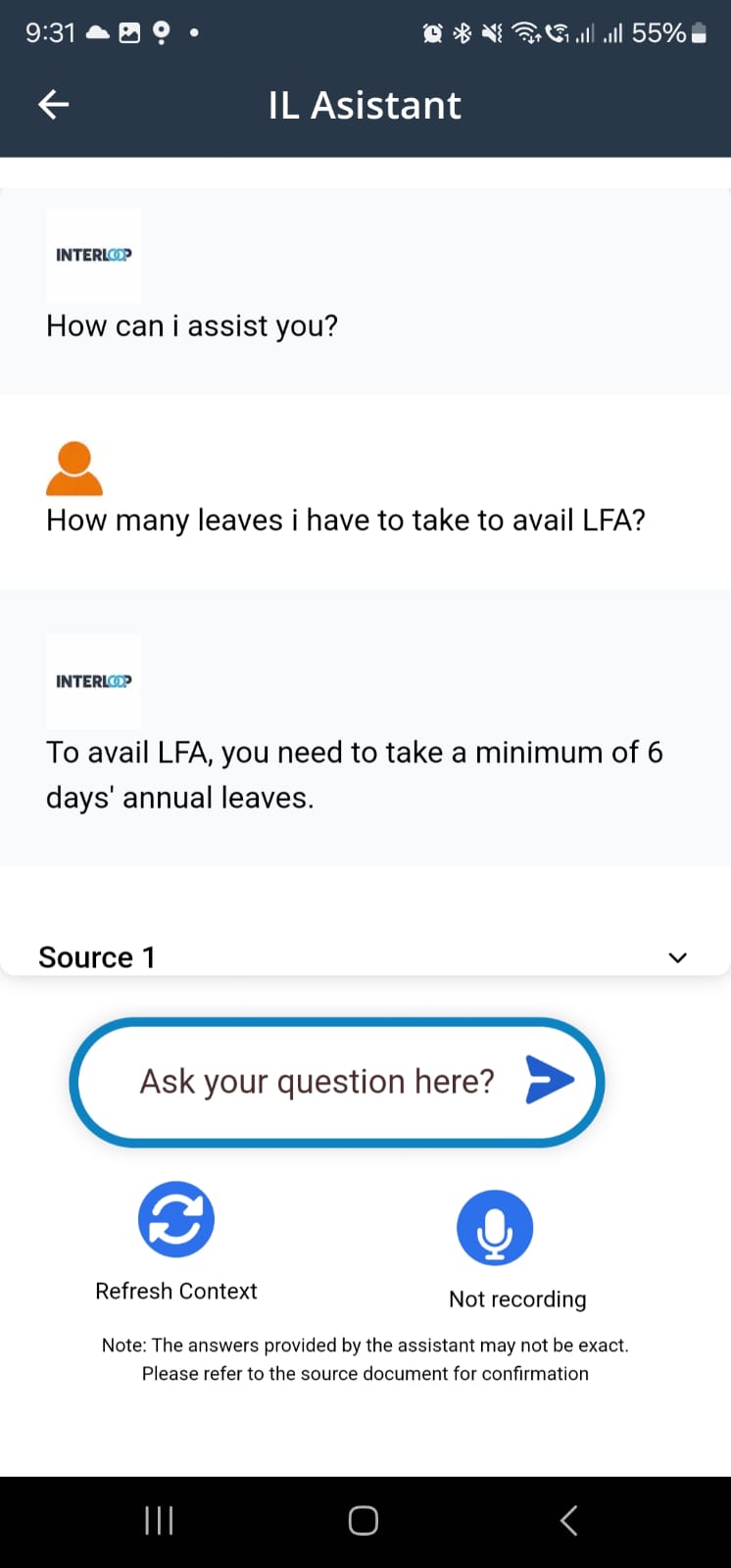} 
    \caption{Mobile Application View Demonstration}
    \label{fig:MobileApp}
\end{figure} 

\subsection{System Performance}

The retrieval augmented generation system has demonstrated robust performance in handling diverse queries across multiple languages and cultural contexts. Through careful integration of large language models, efficient prompt engineering, and strategic delivery setups, the system has proven to be accurate and effective in generating contextually relevant responses.

\begin{figure}
    \centering
    \includegraphics[width=0.2\textwidth]{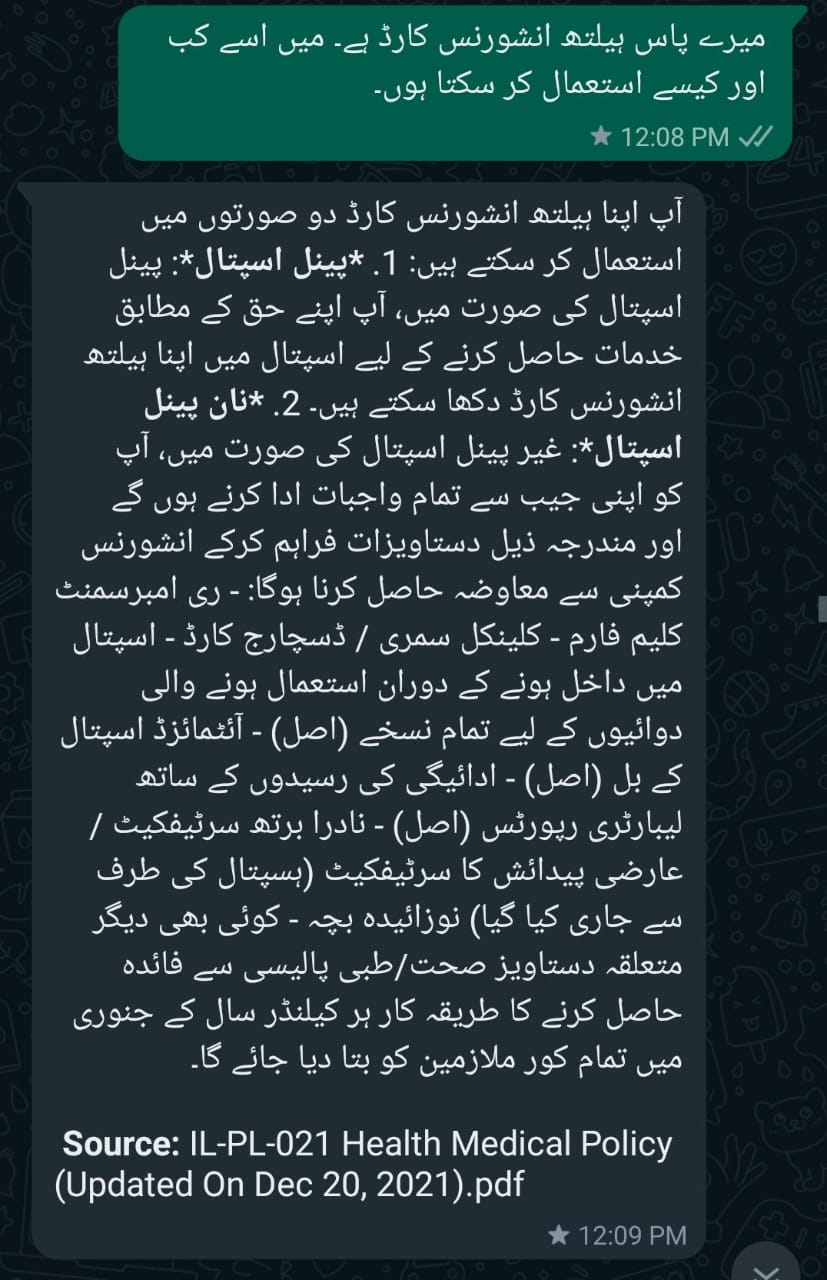} 
    \caption{Whatsapp Demonstration of Urdu Conversation from English Documents}
    \label{fig:WhatsappUrdu}
\end{figure}

\subsection{Impact on HR Queries}

One of the notable achievements of the system is the substantial reduction in HR-related queries. Over the past months, the number of queries received by HR has seen a consistent decline, with a remarkable 30 percent reduction each month. This reduction can be attributed to the system's ability to provide timely and accurate information, empowering users to find answers independently.

\subsection{User Engagement Metrics}

The adoption of the retrieval augmented generation system is reflected in the daily user engagement metrics. On average, the mobile application receives 700 conversations per day, showcasing the active participation and reliance on the system for information retrieval. Additionally, the WhatsApp integration has garnered 450 conversations daily, further extending the reach and accessibility of the system.

\begin{figure}
    \centering
    \includegraphics[width=0.2\textwidth]{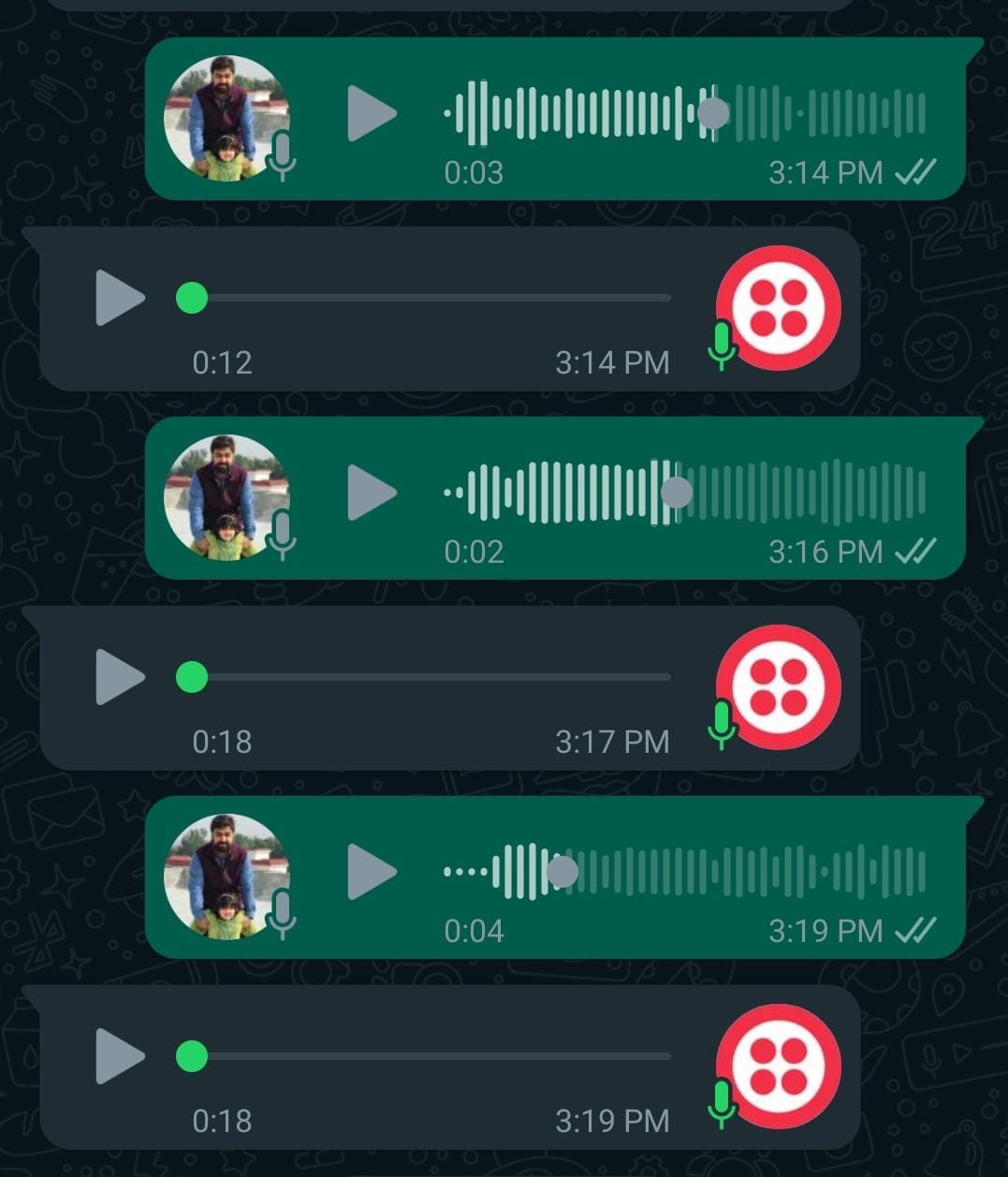} 
    \caption{Quick Responses Over an Audio Message conversation on Whatsapp}
    \label{fig:Audio}
\end{figure} 

\subsection{Complaints}

The efficiency of the system is evident in the low number of complaints received in the last two months. With only 12 complaints reported, the system has demonstrated its reliability and user satisfaction. The limited number of complaints highlights the successful implementation of user-centric design and the system's ability to meet user expectations.

\subsection{Multilingual and Voice Interaction}

The system's adaptability to multilingual communication is evident in the user interactions. Out of a total of 1150 conversations, 45 percent were conducted through voice interactions, showcasing the effectiveness of the implemented speech capability. Moreover, 59 percent of conversations were not conducted in English, underscoring the system's proficiency in handling queries in various languages.

\section{Conclusion}

In conclusion, the retrieval augmented generation system has not only met but exceeded expectations, showcasing exceptional performance in terms of accuracy, user engagement, and cultural adaptability. The positive outcomes observed, including the significant reduction in HR queries, heightened user satisfaction, and successful handling of multilingual interactions, affirm the system's efficacy in enhancing communication and information retrieval within the multicultural enterprise.

Throughout this paper, we successfully devised a sustainable strategy for the adoption of Generative AI in an enterprise environment characterized by a large workforce with varying levels of literacy. The implementation of this strategy has paved the way for improved communication and streamlined information retrieval processes.

However, despite the success achieved, certain challenges remain that need careful consideration in future developments:

\begin{itemize}
    \item \textbf{Optimization of TTS, STT, and LLM Models for Multilingual Calls:} The current system faces challenges in optimizing Text-to-Speech (TTS), Speech-to-Text (STT), and Large Language Model (LLM) models for multilingual calls. Ongoing work on caching LLM responses and audio diffusion models may provide insights and solutions to enhance the efficiency of these components.

    \item \textbf{Cost Effectiveness:} Future work should focus on improving the cost-effectiveness of the system. This involves exploring strategies to optimize resource utilization and potentially adopting more cost-efficient models or technologies.

    \item \textbf{Inclusion of Regional Dialects:} The current speaking capability of the system is limited to English and Urdu responses, excluding Pakistani dialects such as Punjabi, Sindhi, and Balochi. While the system can understand these languages, building custom models for Text-to-Speech to include regional dialects is crucial for enhanced efficiency and effectiveness in communication.
\end{itemize}

Addressing these challenges will further refine and extend the capabilities of the retrieval augmented generation system, ensuring its continued success in serving the diverse needs of a multicultural enterprise.


\bibliographystyle{abbrvnat}


\end{document}